\documentclass[twocolumn,prb,aps,showpacs]{revtex4}
\usepackage{epsf,latexsym,graphicx}

\begin{document}
\newcommand{\av}[1]     {\langle #1 \rangle}
\newcommand{\Av}[1]     {\left\langle #1 \right\rangle}
\newcommand{\comm}[1]   {\left[ #1 \right]}
\newcommand{\antc}[1]   {\left\{ #1 \right\}}
\newcommand{\bra}[1]    {\langle #1 |}
\newcommand{\ket}[1]    {| #1 \rangle}
\newcommand{\eqn}[1]    {(\ref{#1})}
\newcommand{\subbox}[1] {{\mbox{\scriptsize #1}}}
\newcommand{\ex}[1] {{e^{ #1 }}}

\def\etal   {{\em et al.}}

\def\bi         {\begin{itemize}}
\def\ei         {\end{itemize}}
\def\benu   {\begin{enumerate}}
\def\eenu   {\end{enumerate}}
\def\bmat       {\left( \begin{array}}
\def\emat       {\end{array} \right)}
\def\beq    {\begin{equation}}
\def\eeq    {\end{equation}}
\def\beqn       {\begin{eqnarray*}}
\def\eeqn       {\end{eqnarray*}}
\def\beqa       {\begin{eqnarray}}
\def\eeqa       {\end{eqnarray}}
\def\bquote {\begin{quote}}
\def\equote {\end{quote}}
\def\f          {\frac}

\def\a          {\alpha}
\def\b          {\beta}
\def\c          {\chi}
\def\d          {\delta}
\def\e          {\epsilon}
\def\et         {\eta}
\def\g          {\gamma}
\def\k          {\kappa}
\def\l      {\lambda}
\def\m          {\mu}
\def\n          {\nu}
\def\s          {\sigma}
\def\t          {\tau}
\def\th         {\theta}
\def\ve     {\varepsilon}
\def\vph    {\varphi}
\def\w          {\omega}
\def\x          {\xi}
\def\z      {\zeta}

\def\A          {\Alpha}
\def\B          {\Beta}
\def\D          {\Delta}
\def\E          {\Epsilon}
\def\Et         {\Eta}
\def\G          {\Gamma}
\def\tG         {\tilde{\Gamma}}
\def\La          {\Lambda}
\def\M          {\Mu}
\def\Si          {\Sigma}
\def\Th         {\Theta}
\def\W          {\Omega}
\def\X          {\Xi}

\def\ba         {{\bf a}}
\def\bb         {{\bf b}}
\def\bd         {{\bf d}}
\def\bdf        {{\bf f}}
\def\be         {{\bf e}}
\def\bg         {{\bf g}}
\def\bh         {{\bf h}}
\def\bk         {{\bf k}}
\def\bl         {{\bf l}}
\def\bm         {{\bf m}}
\def\bn         {{\bf n}}
\def\bp         {{\bf p}}
\def\bq         {{\bf q}}
\def\br         {{\bf r}}
\def\bs         {{\bf s}}
\def\bv         {{\bf v}}
\def\bx         {{\bf x}}
\def\by         {{\bf y}}
\def\bz         {{\bf z}}
\def\bA         {{\bf A}}
\def\bB         {{\bf B}}
\def\bD         {{\bf D}}
\def\bF         {{\bf F}}
\def\bG         {{\bf G}}
\def\bH         {{\bf H}}
\def\bJ         {{\bf J}}
\def\bM         {{\bf M}}
\def\bQ         {{\bf Q}}
\def\bR         {{\bf R}}
\def\bS         {{\bf S}}
\def\bX         {{\bf X}}
\def\bal        {{\mbox{\boldmath$\a$}}}
\def\bsi    {{\mbox{\boldmath$\s$}}}
\def\bDel       {{\mbox{\boldmath$\Delta$}}}

\def\cA     {{\mathcal{A}}}
\def\cB         {{\mathcal{B}}}
\def\cD     {{\mathcal{D}}}
\def\cH     {{\mathcal{H}}}
\def\cF     {{\mathcal{F}}}
\def\cG     {{\mathcal{G}}}
\def\cK     {{\mathcal{K}}}
\def\cM         {{\mathcal{M}}}
\def\cN     {{\mathcal{N}}}
\def\cO     {{\mathcal{O}}}
\def\cP     {{\mathcal{P}}}
\def\cR     {{\mathcal{R}}}
\def\cV     {{\mathcal{V}}}

\def\hd     {{\hat{d}}}
\def\hk     {{\hat{k}}}
\def\hs     {{\hat{\s}}}
\def\hbr    {{\hat{\br}}}
\def\hbs    {{\mbox{\boldmath$\hat{\s}$}}}
\def\hbx    {{\hat{\bx}}}
\def\hby    {{\hat{\by}}}
\def\hbz    {{\hat{\bz}}}
\def\hps    {{\hat{\psi}}}
\def\hu     {{\hat{u}}}
\def\hv     {{\hat{v}}}
\def\hA     {{\hat{A}}}
\def\hC     {{\hat{C}}}
\def\hD     {{\hat{D}}}
\def\hG     {{\hat{G}}}
\def\hF     {{\hat{F}}}
\def\hH     {{\hat{H}}}
\def\hV     {{\hat{V}}}
\def\hO     {{\hat{O}}}
\def\hU     {{\hat{U}}}
\def\hcD    {{\hat{\mathcal{D}}}}
\def\hcF    {{\hat{\mathcal{F}}}}
\def\hcG    {{\hat{\mathcal{G}}}}
\def\hDel   {{\hat{\D}}}
\def\hSig   {{\hat{\Si}}}
\def\hGu    {{\hat{\underline{G}}}}
\def\hCk    {{\hat{C}_\bk^{}}}
\def\hCkd   {{\hat{C}_\bk^\dagger}}
\def\hCa    {{\hat{C}_\alpha^{}}}
\def\hCad   {{\hat{C}_\alpha^\dagger}}
\def\hCb    {{\hat{C}_\beta^{}}}
\def\hCbd   {{\hat{C}_\beta^\dagger}}
\def\ho     {{\hat{1}}}

\def\tila   {{\tilde{a}}}
\def\tilal  {{\tilde{\a}}}
\def\tilet  {{\tilde{\et}}}
\def\tilG   {{\tilde{G}}}
\def\tilK   {{\tilde{K}}}
\def\tilcG  {{\tilde{\cG}}}
\def\tilw   {{\tilde{\w}}}
\def\tilT   {{\tilde{T}}}
\def\tg     {{\tilde{\g}}}
\def\tD     {{\tilde\Delta}}
\def\bn     {{{\bf\nabla}}}
\def\tR     {{\tilde{R}}}
\def\trho     {{\tilde{\rho}}}

\def\ckD    {{\check{D}}}
\def\ckg    {{\check{g}}}
\def\ckSi   {{\check{\Sigma}}}
\def\oo     {{\otimes}}
\def\barK   {{\bar{K}}}
\def\barcG  {{\bar{\cG}}}

\def\vA     {{\vec{A}}}
\def\vB     {{\vec{B}}}
\def\vM     {{\vec{M}}}
\def\vk     {{\mathbf{k}}}
\def\kp     {{k_{\perp}}}

\def\la     {\langle}
\def\ra     {\rangle}
\def\larr   {\leftarrow}
\def\rarr   {\rightarrow}
\def\llarr  {\longleftarrow}
\def\Llarr  {\Longleftarrow}
\def\lrarr  {\longrightarrow}
\def\Lrarr  {\Longrightarrow}
\def\uparr  {\uparrow}
\def\dnarr  {\downarrow}
\def\dag    {\dagger}

\def\tr     {{\mbox{Tr}~}}
\def\im     {{\mbox{Im}}}
\def\re     {{\mbox{Re}}}
\def\Res    {{\mbox{Res}}}
\def\sgn    {{\mbox{sgn}~}}

\def\cm     {{\mbox{cm}}}
\def\kOe    {{\mbox{kOe}}}

\def\mca        {{M_{\mbox{\scriptsize AF}}}}
\def\mis        {{M_{\mbox{\scriptsize SP}}}}
\def\avmca      {{\bar{M}_{\mbox{\scriptsize AF}}}}
\def\avmis      {{\bar{M}_{\mbox{\scriptsize SP}}}}
\def\aca        {{\a_{\mbox{\scriptsize AF}}}}
\def\ais        {{\a_{\mbox{\scriptsize SP}}}}
\def\bca        {{\b_{\mbox{\scriptsize AF}}}}
\def\bis        {{\b_{\mbox{\scriptsize SP}}}}
\def\bas        {{\b_{\mbox{\scriptsize AF-SP}}}}
\def\ani        {{a_{\mbox{\scriptsize A}}}}
\def\bni        {{b_{\mbox{\scriptsize A}}}}
\def\abr        {{a_{\mbox{\scriptsize B}}}}
\def\bbr        {{b_{\mbox{\scriptsize B}}}}
\def\vphb       {{\vph_{\mbox{\scriptsize B}}}}

\def\vphni      {{\vph_{\mbox{\scriptsize A}}}}
\def\vphb       {{\vph_{\mbox{\scriptsize B}}}}
\def\mni        {{m_{\mbox{\scriptsize A}}}}
\def\mb         {{m_{\mbox{\scriptsize B}}}}
\def\xini       {{\xi_{\mbox{\scriptsize A}}}}
\def\xibr       {{\xi_{\mbox{\scriptsize B}}}}
\def\fni        {{f_{\mbox{\scriptsize A}}}}
\def\fbr        {{f_{\mbox{\scriptsize B}}}}

\def\Ck     {{C_\bk^{}}}
\def\Ckd    {{C_\bk^\dagger}}
\def\cks    {{c_{k {\sigma}}}}
\def\cku    {{c_{k {\uparrow}}}}
\def\ckd    {{c_{k {\downarrow}}}}
\def\cksd    {{c_{k {\sigma}}^\dagger}}
\def\ckud    {{c_{k {\uparrow}}^\dagger}}
\def\ckdd    {{c_{k {\downarrow}}^\dagger}}
\def\cmkdd    {{c_{{-k} {\downarrow}}^\dagger}}
\def\cmkd    {{c_{{-k} {\downarrow}}}}
\def\cmkpd    {{c_{{-k'} {\downarrow}}}}
\def\ckpu    {{c_{{k'} {\uparrow}}}}
\def\aks    {{a_{k {\sigma}}}}
\def\aku    {{a_{k {\uparrow}}}}
\def\akd    {{a_{k {\downarrow}}}}
\def\aksd    {{a_{k {\sigma}}^\dagger}}
\def\akud    {{a_{k {\uparrow}}^\dagger}}
\def\akdd    {{a_{k {\downarrow}}^\dagger}}

\def\uup    {{u_{{\uparrow}}}}
\def\udo   {{u_{{\downarrow}}}}
\def\vup    {{v_{{\uparrow}}}}
\def\vdo    {{v_{{\downarrow}}}}

\def\brag   {{\langle|}}
\def\ketg   {{|\rangle}}

\def\plx     {\f{\partial}{\partial x}}
\def\uu      {\uparrow\uparrow}
\def\ud      {\uparrow\downarrow}
\def\du      {\downarrow\uparrow}
\def\dd      {\downarrow\downarrow}

\def \ba {\begin{eqnarray}}
\def \ea {\end{eqnarray}}
\def \vk {\mathbf{k}}

\def \Bi2212 {Bi$_2$Sr$_2$CaCu$_2$O$_{8+\delta}$}

\title{Analysis of Laser ARPES from \Bi2212 \ in superconductive state:
angle resolved self-energy and fluctuation spectrum }

\author{Jae Hyun Yun}\affiliation{Department of Physics and Institute for
Basic Science Research, SungKyunKwan University, Suwon 440-746,
Korea.}

\author{Jin Mo Bok }\affiliation{Department of Physics and Institute for
Basic Science Research, SungKyunKwan University, Suwon 440-746,
Korea.}

\author{Han-Yong Choi}
\affiliation{Department of Physics and Institute for Basic Science
Research, SungKyunKwan University, Suwon 440-746, Korea. \\
School of Physics, Korea Institute for Advanced Study, Seoul
130-722, Korea. \\
Asia Pacific Center for Theoretical Physics, Pohang 790-784,
Korea.}

\author{Wentao Zhang}
\affiliation{National Laboratory for Superconductivity, Beijing
National Laboratory for Condensed Matter Physics, Institute of
Physics, Chinese Academy of Sciences, Beijing 100190, China.}

\author{X. J. Zhou}
\affiliation{National Laboratory for Superconductivity, Beijing
National Laboratory for Condensed Matter Physics, Institute of
Physics, Chinese Academy of Sciences, Beijing 100190, China.}

\author{Chandra M. Varma}
\affiliation{Department of Physics and Astronomy, University of
California, Riverside, California 92521. }

\begin{abstract}

We analyze the ultra high resolution laser angle resolved
photo-emission spectroscopy (ARPES) intensity from the slightly
underdoped \Bi2212 \ in the superconductive (SC) state. The
momentum distribution curves (MDC) were fitted at each energy $\w$
employing the SC Green's function along several cuts
perpendicular to the Fermi surface with the tilt angle $\theta$
with respect to the nodal cut. The clear observation of
particle-hole mixing was utilized such that the complex
self-energy as a function of $\omega$ is directly obtained from
the fitting. The obtained angle resolved self-energy is then used
to deduce the Eliashberg function $\alpha^2 F^{(+)}(\th,\w)$ in
the diagonal channel by inverting the $d$-wave Eliashberg equation
using the maximum entropy method. Besides a broad featureless
spectrum up to the cutoff energy $\omega_c$, the deduced
$\alpha^2 F$ exhibits two peaks around 0.05 eV and 0.015 eV. The
former and the broad feature are already present in the normal
state, while the latter emerges only below $T_c$. Both peaks
become enhanced as $T$ is lowered or the angle $\th$ moves away
from the nodal direction. The implication of these findings are
discussed.

\end{abstract}
\maketitle


\section{Introduction}

The recent observation of the particle-hole mixing in the
superconductive (SC) state of the cuprates by high resolution
angle-resolved photo-emission spectroscopy (ARPES) has opened up a
new window to probe the fundamental physics of high temperature
superconductivity.\cite{Wentao11prl,Matsui03prl} In particular, an
analysis of the spectra in the SC state, using the Eliashberg
formalism for $d$-wave superconductivity, provides the fluctuation
spectrum responsible for pairing. This is an extension of the
tunneling experiments and analysis with which it was definitively
established that the pairing in metals like Pb is through
exchange of phonons.\cite{McMillan65prl} It should be remembered
that to get reliable information, it was necessary to have
measurements of conductance at different temperatures and range
of voltages of the order of the cut-off energy in the phonon
spectrum to an accuracy of 0.2 \%. The particle-hole mixing in
cuprate superconductors was first observed some 15 years ago in
ARPES.\cite{Campuzano96prb} Those experiments had much worse
momentum and energy resolutions. Since the cut-off is an order of
magnitude higher  for the cuprates than Pb and the
angle-dependence of the spectra is crucial, the demands on the
quality of the data are only being recently met through
ultra-high resolution and stability of laser based ARPES.

The ARPES provided an early evidence for the $d_{x^2-y^2}$ pairing
state of the cuprates.\cite{Shen93prl} The measured leading edge
shift of the energy distribution curve (EDC) of ARPES as a
function of the tilt angle showed that the superconducting gap is
consistent with the $d$-wave pairing gap. It ushered in more
debates and experiments which eventually led to the establishment
of the $d$-wave pairing symmetry for the cuprate
superconductors.\cite{Wollman93prl} The ARPES contains more
information than the leading edge shift which may be utilized, for
example, to extract the Eliashberg functions and track their
evolution as the temperature is lowered below $T_c$. By properly
extending the normal state analysis of extracting the
self-energy, one should be able to deduce information about
superconductivity of the cuprates such as the angle-resolved
diagonal and off-diagonal self-energies, and the pertinent
Eliashberg functions. This is precisely what we wish to present
in this paper.

For this, we fitted the ARPES momentum distribution curves (MDC)
at each energy $\omega$ employing the SC Green's function along
several cuts perpendicular to the Fermi surface with the tilt
angle $\theta$ from the nodal cut with respect to the $(\pi,\pi)$
in the Brillouin zone. The clear observation of particle-hole
mixing was utilized such that the complex self-energy as a
function of $\omega$ is directly obtained from fitting the ARPES
data. Thus obtained angle resolved diagonal self-energy
$\Sigma(\th,\w)$ is then used to deduce the Eliashberg function
$\alpha^2 F^{(+)}(\th,\w)$, i.e., the bosonic fluctuation spectrum
multiplied by the coupling constant squared, in the diagonal
channel by inverting the $d$-wave Eliashberg equation using the
maximum entropy method (MEM). The diagonal self-energy evolves
smoothly into the normal state self-energy as the temperature is
raised above $T_c$. The evolution of the Eliashberg function as
the temperature or tilt angle is varied will reveal a useful
information about the nature of superconductivity in the cuprates.

On the other hand, the angle resolved off-diagonal self-energy
$\phi(\th,\w)$, or, the density of states $N(\th,\w)$ given by
Eq.\ (\ref{returns}) below, can be used for $d$-wave
superconductors to extract the Eliashberg function in the
off-diagonal (i.e., pairing) channel\cite{Vekhter03prl} as the
ordinary tunneling conductance was used by McMillan and Rowell to
extract the spectrum of fluctuations for $s$-wave
superconductors.\cite{McMillan65prl} While the Eliashberg
functions along the diagonal and off-diagonal channels are
assumed to be the same for the $s$-wave pairing, they are in
general different for $d$-wave superconductors. The current
approach has the unique advantage in that it can disentangle the
Eliashberg functions in the diagonal and off-diagonal channels,
$\a^2 F^{(+)}(\th,\w)$ and $\a^2 F^{(-)}(\th,\w)$, respectively.

In the following section II, we will present the formulation of
the MDC analysis of the ARPES intensity in the superconducting
state using the full momentum and energy dependence of SC Green's
function. It is an extension of the ARPES analysis in the normal
state.\cite{Bok10prb} The results for the diagonal self-energy
$\Sigma(\th,\w)$ from slightly underdoped Bi2212 will be
presented in section III at temperatures above and below $T_c$
and along several cuts of the tilt angle $\theta$ with respect to
the $(0,0)-(\pi,\pi)$ nodal cut. As in the normal state, the
extracted self-energy may be used as an input to deduce the
Eliashberg function $\alpha^2 F^{(+)} (\theta,\omega)$. The
obtained Eliashberg functions are presented in section IV. Recall
that $\alpha^2 F^{(+)}(\theta,\omega)$ along different cuts
collapse onto a single curve with a peak near 0.05 eV below the
angle dependent cutoff $\omega_c(\theta)$ in the normal
state.\cite{Bok10prb} In the SC state the peak around 0.05 eV
gets enhanced and additional peak emerges around 0.015 eV below
$T_c$. Both peaks become enhanced as $T$ is lowered or the tilt
angle is increased. We will conclude the paper by making some
remarks and outlooks in the section V.


\section{Formalism}

\begin{figure}
\includegraphics[width=3.1in]{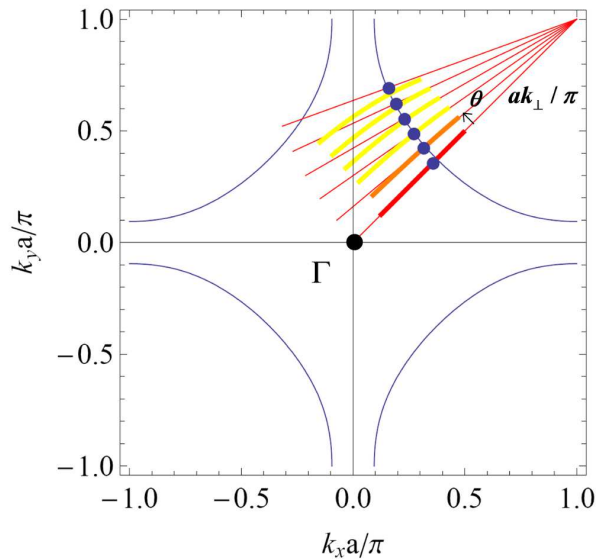}
\caption{The Fermi surface of Bi2212 in the Brillouin zone. The
blue solid curves centered around the $\Gamma$ point is the FS
from Eq.\ (4\ref{TB}) and the solid dots are the experimentally
determined FS at $\theta=0^{\circ}, 5^{\circ}, 10^{\circ},
15^{\circ}, 20^{\circ}, 25^{\circ}$. $\kp$ is the distance from
the $(\pi,\pi)$ point. The yellow thick curves along each cut
indicate the actual momentum paths at $\w=0$ of the experimentally
measured ARPES MDC data.} \label{Fermisurface}
\end{figure}

The ARPES intensity, within the sudden approximation, is given by
 \ba
I(\vk,\w)= |M(\vk,\nu)|^2 f(\w) \left[A(\vk,\w)+B(\vk,\w)\right],
 \ea
where $M(\vk,\n)$ is the matrix element, $\n$ the energy of
incident photon, $f(\w)$ the Fermi distribution function,
$A(\vk,\w)$ the quasiparticle (qp) spectral function, and
$B(\vk,\w)$ is the background from the scattering of the
photo-electrons. We write the in-plane momentum $\vk$ with the
distance from the $(\pi,\pi)$ point $k_\perp$ and the tilt angle
measured from the nodal cut $\theta$ as shown in Fig.\
\ref{Fermisurface}. The self-energy has a much weaker dependence
on $k_\perp$ than $\theta$ or $\omega$ as will be discussed
below. Assuming this, the spectral function is written as
 \beqa
A(\vk,\w)=-\f{1}{\pi}\textrm{Im}G(\vk,\w),\nonumber\\
G(\vk,\w)=\f{W(\theta,\w)+Y(\vk,\w)}{W^{2}(\theta,\w)-Y^2(\vk,\w)-\phi^2(\theta,\w)},
 \eeqa
where $G(\vk,\w)$ is the retarded Green's function in the
superconductive state. The following notations were used:
 \beqa
W(\theta,\w)=\w Z(\theta,\w)=\w-\Si(\theta,\w), \nonumber \\
Y(\vk,\w)= \xi(\vk)+X(\th,\w), \nonumber \\
 \phi(\theta,\w)= Z(\theta,\w) \Delta(\theta,\w),
 \eeqa
where $Z(\theta,\w)$ is the renormalization function, $X(\th,\w)$
the shift of the qp dispersion, and $\Si(\theta,\w)$ and
$\phi(\theta,\w)$ represent the qp diagonal self-energy and the
off-diagonal self-energy, respectively. The equation that
connects $\Si(\theta,\w)$, $X(\th,\w)$, and $\phi(\th,\w)$ with
the effective interaction in the charge and spin channels is the
Eliashberg equation.\cite{Sandvik04prb} It is presented in section
IV below in connection with extraction of the Eliashberg
functions.

It is informative to make the following decomposition of the SC
Green's function:
 \beqa\label{decompo}
\f{Y+W}{Y^{2}-\left(W^2-\phi^2 \right)} = \frac{1/2+N/2}{Y-P} +
\frac{1/2-N/2}{Y+P},
 \eeqa
where
 \beqa
P(\th,\w)=\sqrt{W^{2}( \th,\w)-\phi^{2}(\th,\w)},
 \\ N(\th,\w)=
\frac{W(\th,\w)} {\sqrt{W^{2}(\th,\w)-\phi^{2}(\th,\w)} }.
\label{returns}
 \eeqa
We note that the qp dispersion shift $X(\th,\w)$ vanishes in the
particle-hole symmetric band. Although the symmetry does not hold
for the realistic tight-binding dispersion, it holds to a good
degree over the small energy scale of SC and the renormalization
is neglected in the present work. The ARPES intensity devided by
the Fermi distribution function is then given by
\begin{widetext}
 \beqa\label{fiteq}
\f{I(\th,\kp,\w)}{f(\w)}=C(\th,\w) \textrm{Im}\left[\f
{1+N(\th,\w)}{\x(\vk)-P(\th,\w)}+\f
{1-N(\th,\w)}{\x(\vk)+P(\th,\w)}\right]+B(\th,\w),
 \eeqa
\end{widetext}
where $C(\th,\w)$ is the weight of the spectral function of the
ARPES intensity. We then have the six parameter fit in SC state:
$C$, $B$, the real and imaginary parts of $P$ and $N$ as a
function of binding energy $\w$, while the normal state fitting
required four parameters. Note that the dependence on $k_\perp$
comes in through the bare dispersion $\x(\vk)$ only. It is
therefore important to take appropriate dispersion.

As in the normal state, we used the tight-binding (TB) dispersion
and the linear dispersion (LD) for the MDC analysis. The TB
dispersion $\x(\vk)$ is given by
 \beqa\label{TB}
\x(k_{x},k_{y})=-2t(\cos k_{x}a+\cos k_{y}a)+4t'\cos k_{x}a\cos
k_{y}a\nonumber\\-2t''(\cos2k_{x}a+\cos 2k_{y}a)-\m,
 \eeqa
where $a=3.82$ {\AA} is the lattice constant and $\m$ is the
chemical potential. We took $t=0.395$, $t'=0.084$, $t''=0.042$,
and $\m=-0.43$ eV. The linear dispersion was determined by
linearization of the TB at FS of the six tilt angles $\th$.
 \beqa
\x(\th,\kp)=v_{F}(\th)\left[\kp-k_{F}(\th)\right],
 \eeqa
where $v_{F}(\th)$ and $k_{F}(\th)$ are Fermi velocity and Fermi
momentum, respectively. The experimentally determined FS in
comparison with that from Eq.\ (\ref{TB}) is shown in Fig.\
\ref{Fermisurface}. The six cuts with the tilt angles $\th$ with
respect to the $(\pi,\pi)$ are also shown with the solid lines.

Note that the first and the second terms in Eq. (\ref{fiteq})
give the intensity due to the ``particle'' and the ``hole'' parts
of the Bogoliubov particles, respectively. $P(\th,\w)$ and
$N(\th,\w)$ are directly extracted from fitting the ARPES MDC
data. Then,
 \ba
\Sigma(\theta,\w)=\w-P(\theta,\w)N(\theta,\w), \label{sig} \\
\phi^2(\theta,\w) = P^2(\theta,\w)[ N^2(\theta,\w) -1 ]\label{phi}
 \ea
gives $\Sigma(\theta,\w)$ and $\phi(\theta,\w)$. Since the
density of states $N(\th,\w) =1$ in the normal state, the
off-diagonal self-energy $\phi(\theta,\w)$ can only be extracted
from the difference of the spectra between the normal and SC
state. These differences are very small at energies above a few
times $T_c$. So the requirements on the ARPES data to reliably
extract $\phi$ at higher energies are considerably more stringent
than those to extract $\Sigma$. We defer this to future work and
show here that considerable information on the fluctuation
spectrum can be extracted from the diagonal self-energy alone.

\begin{figure}
\includegraphics[width=3.8in]{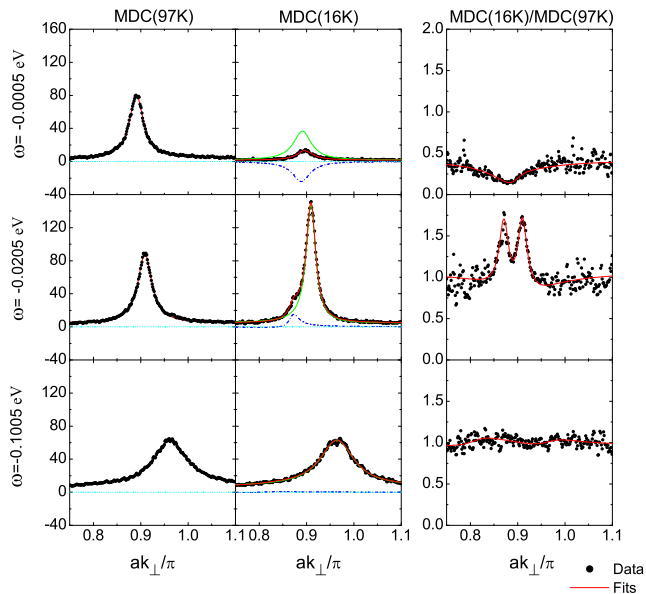}
\vspace{-0.1in}
 \caption{ The representative MDC as a function of
the momentum along the tilt angle $\th=20^{\circ}$. The dots are
the experimental data and the solid red lines are the fitting. The
first and second columns show the fittings in the normal state
and in the SC state, respectively. The last column is the MDC
ratios of SC to normal states.}\label{MDCfitSC}
\end{figure}

\section{The ~MDC ~analysis}

The ultra high resolution Laser ARPES data were collected from
slightly underdoped Bi2212 of SC critical temperature $T_c=89$ K
and pseudogap temperature $T^*\approx 160$ K. The data were took
along the cuts of the tilt angle $\th=0$ (nodal cut), 5, 10, 15,
20, and 25 degrees with respect to the nodal direction and at
temperatures $T=107,$ 97 above $T_c$ and 80, 70, and 16 K below
$T_c$. The photon energy of $h\n=6.994$ eV was used in the laser
ARPES. Refer to Ref.\ \cite {Bok10prb} for more detailed
description of the experimental setup and the technical details.

Typical results of the ARPES analysis in SC state are given in
Fig.\ \ref{MDCfitSC} for the tilt angle $\th=20^{\circ}$. The
first column shows the results in the normal state at $T=97$ K as
a function of the magnitude of the inplane momentum $k_\perp$
from the $(\pi,\pi)$ point at the energy $\w = -0.0005,~
-0.0205$, and $-$0.1005 eV. The three energies represent the
cases of $\w \ll \Delta$, $\w\approx\Delta$, and $\w \gg \Delta$,
where $\Delta$ is the gap amplitude at $\th=20^\circ$ and $T=16$
K,
 \ba
\Delta = \frac{\phi(\th,\Delta)}{Z(\th,\Delta)}.
 \ea
The symbols are the data and the red solid lines are the fitting
results. The agreements are almost perfect which justifies the
neglect of the $k_\perp$ dependence of the self-energy. The
second column is the corresponding results deep in the SC state
at $T$ = 16 K. The green solid (blue dashed) lines are the
particle (hole) branch of the fitting, the first (second) term of
Eq.\ (\ref{fiteq}).

The important point is that the hole branch represented by the
blue curves exhibits a  peak as a consequence of the
particle-hole mixing of the pairing. This can be most
spectacularly seen near $\w\approx\Delta$ presented in the middle
row. In addition to the main peak near $k_\perp a/\pi\approx 0.91
$ from the original qp branch, there exists the secondary peak at
$k_\perp a/\pi\approx 0.87 $. This is a direct observation of the
particle-hole mixing deep in the SC state. The details of
observations are presented separately.\cite{Wentao11prl} The
particle-hole mixing was previously reported in the EDC by
observing the bending-back of the spectral
peaks.\cite{Campuzano96prb} Both branches of the Bogoliubov
dispersion due to the particle-hole mixing were also reported by
the EDC in the intermediate temperature regime\cite{Matsui03prl}
because in the low temperature limit the Fermi function cuts the
hole branch off and close to $T_c$ the pairing feature is very
weak. The mixing is observed in the low temperature regime here
and will be utilized to obtain information about
superconductivity in the cuprates. The last row shows the case of
$\w\gg\Delta$. As the energy increases above $\Delta$, the hole
branch contribution vanishes as the bottom plots show. The last
column is the ratios of the MDC at 16 K to 97 K which show the
hole branch more clearly.

\begin{figure}
\includegraphics[width=3.8in]{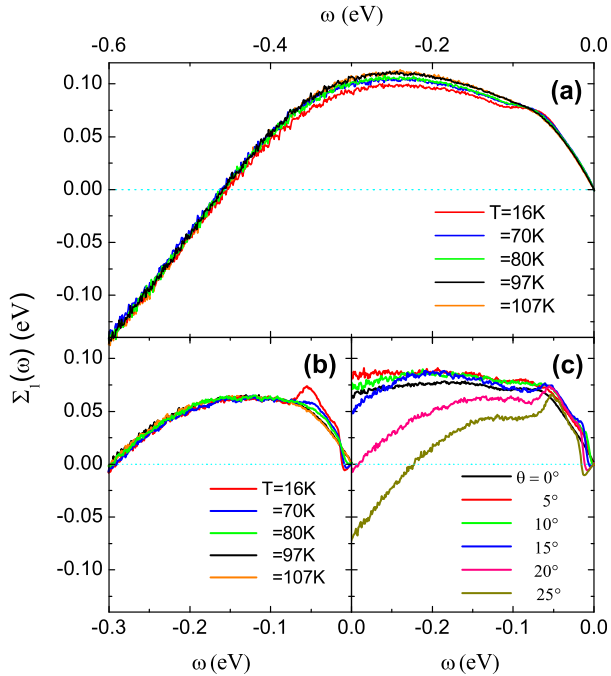}
\vspace{-0.1in}
 \caption{The real part of the extracted
self-energy. The plots (a) and (b) are along the nodal cut and
$\th=20^\circ$, respectively, and the plot (c) is at temperature
$T=16$ K. }\label{sigma16K}
\end{figure}

We now show the real part of the extracted self-energy along the
tilt angle $\th=0$ and $\th=20^\circ$ in Fig.\ \ref{sigma16K}(a)
and (b), respectively, and that at $T=16$ K in the plot (c). Fig.\
\ref{sigma16K}(a) and \ref{sigma16K}(b) demonstrate that the
feature around $0.05-0.07$ eV is already present in the normal
state and is enhanced as the temperature is lowered, while the
broad feature continues from the normal to SC state with no
discernible change within the accuracy of the experiment. The
$0-0.02$ eV feature emerges only along off-nodal cuts below $T_c$
as can be seen from the plots (a) and (b), and its energy scale
increases as the tilt angle is increased as can be seen from the
plot (c). This is consistent with the $d$-wave pairing gap and
implies that the $0-0.02$ eV feature is induced by
superconductivity.

\section{The~ Eliashberg~ function}

The extracted diagonal and off-diagonal self-energies may be used
as experimental inputs to deduce the Eliashberg functions
$\alpha^2 F^{(+)} (\th,\w)$ and $\alpha^2 F^{(-)} (\th,\w)$ by
inverting the Eliashberg equation. The $d$-wave Eliashberg
equation may be written as
 \ba\label{Elidiag}
\Si(\th,\w) = \int^{\infty}_{-\infty} d\e \int^{\infty}_{-\infty}
d\e' S(\w,\e,\e') N_1 (\e) \alpha^2 F^{(+)} (\th,\e'), \\
\phi(\w) = -\int^{\infty}_{-\infty} d\e \int^{\infty}_{-\infty}
d\e' S(\w,\e,\e') D_1 (\e) \alpha^2 F^{(-)}(\e') \label{Elioff},\\
S(\w,\e,\e') = \f{f(\e)+n(-\e')}{\e+\e'-\w-i\d},
 \ea
where $f$ and $n$ represent the Fermi and Bose distribution
functions, respectively.

We took
 \ba
\phi(\th,\w)=\phi(\w)\sin(2\th)
 \ea
because the pairing is $d$-wave, and use the notations
 \ba
N_1 (\e) \equiv \left\langle Re
\frac{W(\th',\e)}{\sqrt{W^2(\th',\e)-\phi^2(\e)\sin^2(2\th')}}
\right\rangle_{\th'}, \\
 D_1 (\e) \equiv \left\langle
\frac{1}{v_{F}(\th')} Re
\frac{\phi(\e)\sin^2(2\th')}{\sqrt{W^2(\th',\e)-\phi^2(\e)\sin^2(2\th')}}
\right\rangle_{\th'}, \\
 \alpha^2 F^{(+)} (\th,\e')\equiv
\left\langle
\f{\a^{2}(\th,\th')}{v_{F}(\th')}F^{(+)}(\th,\th',\e')
\right\rangle_{\th'}, \label{a2f}
 \ea
where $v_{F}(\th')$ is the angle-dependent Fermi velocity and the
bracket implies the angular average over $\th'$.

As in the normal state, we invert the real part of the Eliashberg
equation to deduce the Eliashberg function $\alpha^2
F^{(+)}(\th,\e')$ using the real part of the extracted diagonal
self-energy as an input. As mentioned before, the requirement on
the data and numerical fitting are considerably stringent to
reliably extract the off-diagonal self-energy than to extract the
diagonal self-energy. We will defer the deduction of $\alpha^2
F^{(-)}(\th,\w)$ to future work and focus on $\alpha^2
F^{(+)}(\th,\w)$ here. The real part of Eq.\ (\ref{Elidiag}) may
be written as
 \ba\label{Eliash_real}
\Si_1(\th,\w)=\int^{\infty}_{-\infty} d\w' K(\w,\w') \alpha^2 F^{(+)} (\th,\w'),
 \nonumber \\
K(\w,\w')= \int^{\infty}_{-\infty} d\e \ {\cal P}
\f{f(\e)+n(-\w')}{\e+\w'-\w} N_1 (\e)
 \ea
where ${\cal P}$ represents the principal value, and the
subscripts 1 and 2 refer to the real and imaginary parts. The
inversion was performed using the maximum entropy
method.\cite{numericalrecipes,Bok10prb} Recall that in the normal
state the Eliashberg functions along different cuts of the tilt
angle $\th$ all collapse onto a single curve which has a small
peak at $\approx $ 0.05 eV, flattens above 0.1 eV, and vanishes
above the angle dependent cutoff $\w_c(\th)$.
$\w_c(\th)\approx0.35-0.4$ eV along the nodal direction and
decreases as $\th$ increases.\cite{Bok10prb}

\begin{figure}
\includegraphics[width=3in]{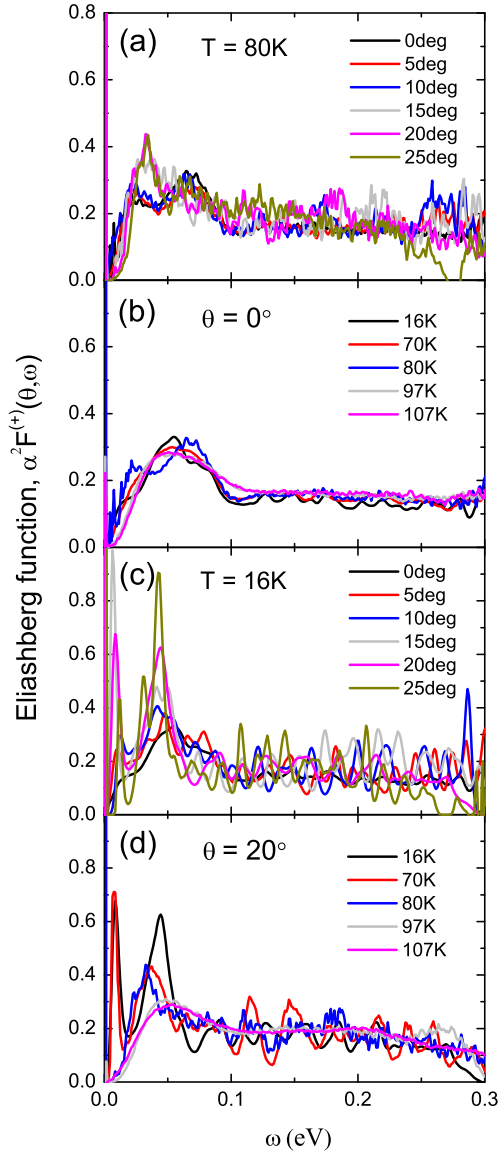}
\vspace{-0.in}
 \caption{The deduced Eliashberg function $\alpha^2
F^{(+)}(\th,\w)$. Plot (a) is at $T=80$ K slightly below $T_c$,
plot (b) along the nodal cut, plot (c) at $T=16$ K deep in the SC
state, and (d) is along the off-nodal cut of $\th=20^\circ$.
}\label{F0}
\end{figure}

Fig.\ \ref{F0}, showing the deduced $\a^2 F^{(+)}(\th,\w)$, is
the key results of the present paper. The deduced function is
noisier at larger angles. The noise somewhat depends on the
multiplier $\alpha$ of the maximum entropy method.\footnote{The
multiplier $\alpha$ is a determinative parameter that controls
how close the fitting should follow the data while not violating
the physical constraints. When $\alpha$ is small, the fitting
will follow the data as closely as possible at the expense of a
noisy and/or negative Eliashberg function, and when $\alpha$ is
large, the extracted Eliashberg function will not deviate much
from the constraint function. } We believe most of the oscillatory
behavior seen are artifacts of the MEM analysis and will focus
only on the robust features whose variation is continuous as a
function of temperature and angles. The broad feature above about
0.07 eV does not change with angle or with temperature. It is a
continuation of that required for the marginal Fermi liquid
properties which were derived recently to arise from quantum
criticality.\cite{Vivek10prb} It carries about 3/4 of the total
spectral weight.

The plot \ref{F0}(a) is slightly below $T_c$ at $T=80$ K. The
Eliashberg function does not change much from the normal state
shape except that the peak value at $\w\approx 0.05$ eV
increases to approximately 0.4 for large tilt angles from 0.3 of
the normal state value. In Fig.\ \ref{F0}(b), $\a^2
F^{(+)}(\th,\w)$ along the nodal cut is shown as the temperature
is varied. As might be expected from above behavior, there is
little change along the nodal cut, although there is a sign of
the lower energy peak at $T=16$ K. It seems that the change in
$\Sigma(\th=0,\w)$ as $T$ is varied as shown in Fig.\
\ref{sigma16K}(a) is predominantly from the change in the density
of states that enters Eq.\ (\ref{Eliash_real}).

$\a^2 F^{(+)}(\th,\w)$ deep in the SC state is shown in Fig.\
\ref{F0}(c). Like at $T=80$ K, the peak at $\w=0.05$ has its
normal state value for small tilt angle and increases as the
angle increases. Also a second peak at $\approx 0.015$ eV emerges
which, like the $\w\approx 0.05$ eV peak, increases from the
normal state value as the angle is increased. In Fig.\
\ref{F0}(d), the Eliashberg function along the cut $\th=20^\circ$
is shown. As the temperature is lowered below $T_c$, the 0.05 eV
peak is enhanced and the 0.015 eV peak newly develops, as one can
anticipate from preceding discussion. Both peaks are enhanced as
$T$ is lowered or the tilt angle is increased.

There have been many investigations of the fluctuation spectrum
of the cuprate superconductors such as the infrared conductivity,
inelastic neutron scattering, Raman scattering, scanning tunneling
spectroscopy, and so on, which are less direct in the information
they provide for the source of superconductivity than ARPES.
Analysis of the frequency dependent conductivity by the McMaster
group reported that a single peak shows up in the Eliashberg
function below 0.1 eV for \Bi2212 \ and other cuprate compounds,
but a double peak feature for La$_{1.83}$Sr$_{0.17}$CuO$_4$ at
low temperatures. It exhibits a peak at $\omega\approx 0.05$ eV
at high temperature $T=250$ K, but as $T$ is lowered to 30 K it
showed two peaks at 0.015 and 0.044 eV.\cite{Hwang08prl} The
inelastic neutron scattering (INS) experiment on
La$_{1.84}$Sr$_{0.16}$CuO$_4$ also reported the two peak structure
around 0.018 and $0.04-0.07$ eV at the antiferromagnetic wave
vector.\cite{Vignolle07naturephys} The Eliashberg analysis of the
break junction SIS conductance and scanning tunneling spectra on
overdoped \Bi2212 \ reported a single peak near $\approx 0.02$
eV.\cite{Zasadzinski06prl}

\begin{figure}
\includegraphics[width=3.7in]{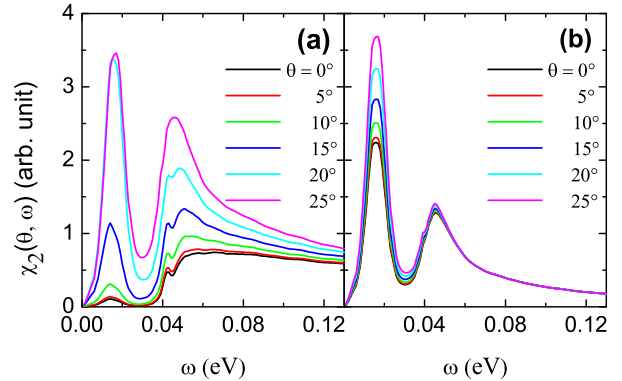}
\vspace{-0.2in}
 \caption{The $\th'$ averaged susceptibility
$\chi_2(\th,\w)$ calculated from Vignolle to compare with $\a^2
F^{(+)}(\th,\w)$. The plot (a) is the $\chi_2(\th,\w)$ from
Vignolle and (b) is with the correlation length reduced to 0.1 of
(a). }\label{spin_sus}
\end{figure}

No inelastic neutron scattering results are available for \Bi2212
, so that we can only compare our results with the detailed
extraction of the spectral function of the magnetic fluctuations
$\chi(\vk,\omega)$ for La$_{1.84}$Sr$_{0.16}$CuO$_4$. The
positions of the two peaks around 0.018 and $0.04-0.07$ eV are
consistent with the peak positions of the deduced $\alpha^2
F^{+}(\theta,\w)$. To make comparison of the momentum dependence
between our $\a^2 F^{(+)}(\th,\w)$ and INS results of
$\chi_2({\bf q},\w)$, we compute $\chi_2(\theta,\omega)$ by
taking the integral over $\theta'$ with both $\vk$ and $\vk'$ on
the Fermi surface having the tilt angles $\th$ and $\th'$,
respectively. As with Eq.\ (\ref{a2f}), we take
 \ba
\chi_2 (\th,\w) \equiv \left\langle \chi_2 (\vk-\vk',\w)
\right\rangle_{\th'}.\label{chiVignolle}
 \ea
The imaginary part of the spin susceptibility $\chi_2 ({\bf
q},\w)$ was taken from Vignolle $et~al.$ INS
results.\cite{Vignolle07naturephys} Fig.\ \ref{spin_sus}(a) is
the computed $\chi_2(\th,\w)$ from Vignolle and (b) is that with
the correlation length reduced to 1/10 of (a). Comparing with our
results in Fig.\ \ref{F0}, the variation with angle compares
better with the reduced correlation length. The angle
independence of the flat part of $ \alpha^2
F^{(+)}(\theta,\omega)$ is consistent only with the fluctuation
spectrum which has the short correlation length on the scale of
the lattice constant.

The origin of the $\sim50$ meV feature is often taken to be a
phonon\cite{Lanzara01nature,Zhou03nature} but it may be the strong
dispersionless magnetic feature recently observed
\cite{Li10nature,He11prl} only in the pseudogap region in the
cuprates. One may be able to decide between the two through ARPES
analysis similar to that done here of overdoped samples without
the pseudogap.

%
%
%

\section{Concluding~remarks}\label{sec:conclusion}

We have presented the analysis of the ARPES intensity in the
superconductive state for the Eliashberg function along the
diagonal channel, $\a^2 F^{(+)}(\th,\w)$. Beside the broad
featureless spectrum, a peak at 0.05 eV present in the normal
state is enhanced as $T$ is lowered, and a second peak emerges
around 0.015 eV in the SC state. The 0.015 eV peak is an
interesting new feature which needs further exploration. Since it
appears only below $T_c$, it can not be responsible for $T_c$. We
do not have an answer at present to the question as to what kind
of collective modes of the superconducting state it represents.
The origin of the 50 meV feature generally observed is at the
heart of current debate. One may be able to answer through ARPES
analysis similar to that done here of overdoped samples without
the pseudogap.

\begin{acknowledgments}

This work was supported by National Research Foundation (NRF) of
Korea through Grant No.\ NRF 2010-0010772. CMV's work is partially
supported by NSF grant DMR-0906530.

\end{acknowledgments}

\bibliographystyle{apsrev}
\bibliography{ref}

\begin{thebibliography}{18}
\expandafter\ifx\csname natexlab\endcsname\relax\def\natexlab#1{#1}\fi
\expandafter\ifx\csname bibnamefont\endcsname\relax
  \def\bibnamefont#1{#1}\fi
\expandafter\ifx\csname bibfnamefont\endcsname\relax
  \def\bibfnamefont#1{#1}\fi
\expandafter\ifx\csname citenamefont\endcsname\relax
  \def\citenamefont#1{#1}\fi
\expandafter\ifx\csname url\endcsname\relax
  \def\url#1{\texttt{#1}}\fi
\expandafter\ifx\csname urlprefix\endcsname\relax\def\urlprefix{URL }\fi
\providecommand{\bibinfo}[2]{#2}
\providecommand{\eprint}[2][]{\url{#2}}

\bibitem[{\citenamefont{Zhang et~al.}(2011)\citenamefont{Zhang, Bok, Yun, Liu,
  Zhao, Liu, Meng, Dong, Zhang, Lu et~al.}}]{Wentao11prl}
\bibinfo{author}{\bibfnamefont{W.}~\bibnamefont{Zhang}},
  \bibinfo{author}{\bibfnamefont{J.~M.} \bibnamefont{Bok}},
  \bibinfo{author}{\bibfnamefont{J.~H.} \bibnamefont{Yun}},
  \bibinfo{author}{\bibfnamefont{G.}~\bibnamefont{Liu}},
  \bibinfo{author}{\bibfnamefont{L.}~\bibnamefont{Zhao}},
  \bibinfo{author}{\bibfnamefont{H.}~\bibnamefont{Liu}},
  \bibinfo{author}{\bibfnamefont{J.}~\bibnamefont{Meng}},
  \bibinfo{author}{\bibfnamefont{X.}~\bibnamefont{Dong}},
  \bibinfo{author}{\bibfnamefont{J.}~\bibnamefont{Zhang}},
  \bibinfo{author}{\bibfnamefont{W.}~\bibnamefont{Lu}}, \bibnamefont{et~al.},
  \bibinfo{journal}{submitted to Phys. Rev. Lett.}  (\bibinfo{year}{2011}).

\bibitem[{\citenamefont{Matsui et~al.}(2003)\citenamefont{Matsui, Sato,
  Takahashi, Wang, Yang, Ding, Fujii, Watanabe, and Matsuda}}]{Matsui03prl}
\bibinfo{author}{\bibfnamefont{H.}~\bibnamefont{Matsui}},
  \bibinfo{author}{\bibfnamefont{T.}~\bibnamefont{Sato}},
  \bibinfo{author}{\bibfnamefont{T.}~\bibnamefont{Takahashi}},
  \bibinfo{author}{\bibfnamefont{S.-C.} \bibnamefont{Wang}},
  \bibinfo{author}{\bibfnamefont{H.-B.} \bibnamefont{Yang}},
  \bibinfo{author}{\bibfnamefont{H.}~\bibnamefont{Ding}},
  \bibinfo{author}{\bibfnamefont{T.}~\bibnamefont{Fujii}},
  \bibinfo{author}{\bibfnamefont{T.}~\bibnamefont{Watanabe}}, \bibnamefont{and}
  \bibinfo{author}{\bibfnamefont{A.}~\bibnamefont{Matsuda}},
  \bibinfo{journal}{Phys. Rev. Lett.} \textbf{\bibinfo{volume}{90}},
  \bibinfo{pages}{217002} (\bibinfo{year}{2003}).

\bibitem[{\citenamefont{McMillan and Rowell}(1965)}]{McMillan65prl}
\bibinfo{author}{\bibfnamefont{W.~L.} \bibnamefont{McMillan}} \bibnamefont{and}
  \bibinfo{author}{\bibfnamefont{J.~M.} \bibnamefont{Rowell}},
  \bibinfo{journal}{Phys. Rev. Lett.} \textbf{\bibinfo{volume}{14}},
  \bibinfo{pages}{108} (\bibinfo{year}{1965}).

\bibitem[{\citenamefont{Campuzano et~al.}(1996)\citenamefont{Campuzano, Ding,
  Norman, Randeria, Bellman, Yokoya, Takahashi, Katayama-Yoshida, Mochiku, and
  Kadowaki}}]{Campuzano96prb}
\bibinfo{author}{\bibfnamefont{J.~C.} \bibnamefont{Campuzano}},
  \bibinfo{author}{\bibfnamefont{H.}~\bibnamefont{Ding}},
  \bibinfo{author}{\bibfnamefont{M.~R.} \bibnamefont{Norman}},
  \bibinfo{author}{\bibfnamefont{M.}~\bibnamefont{Randeria}},
  \bibinfo{author}{\bibfnamefont{A.~F.} \bibnamefont{Bellman}},
  \bibinfo{author}{\bibfnamefont{T.}~\bibnamefont{Yokoya}},
  \bibinfo{author}{\bibfnamefont{T.}~\bibnamefont{Takahashi}},
  \bibinfo{author}{\bibfnamefont{H.}~\bibnamefont{Katayama-Yoshida}},
  \bibinfo{author}{\bibfnamefont{T.}~\bibnamefont{Mochiku}}, \bibnamefont{and}
  \bibinfo{author}{\bibfnamefont{K.}~\bibnamefont{Kadowaki}},
  \bibinfo{journal}{Phys. Rev. B} \textbf{\bibinfo{volume}{53}},
  \bibinfo{pages}{R14737} (\bibinfo{year}{1996}).

\bibitem[{\citenamefont{Shen et~al.}(1993)\citenamefont{Shen, Dessau, Wells,
  King, Spicer, Arko, Marshall, Lombardo, Kapitulnik, Dickinson
  et~al.}}]{Shen93prl}
\bibinfo{author}{\bibfnamefont{Z.-X.} \bibnamefont{Shen}},
  \bibinfo{author}{\bibfnamefont{D.~S.} \bibnamefont{Dessau}},
  \bibinfo{author}{\bibfnamefont{B.~O.} \bibnamefont{Wells}},
  \bibinfo{author}{\bibfnamefont{D.~M.} \bibnamefont{King}},
  \bibinfo{author}{\bibfnamefont{W.~E.} \bibnamefont{Spicer}},
  \bibinfo{author}{\bibfnamefont{A.~J.} \bibnamefont{Arko}},
  \bibinfo{author}{\bibfnamefont{D.}~\bibnamefont{Marshall}},
  \bibinfo{author}{\bibfnamefont{L.~W.} \bibnamefont{Lombardo}},
  \bibinfo{author}{\bibfnamefont{A.}~\bibnamefont{Kapitulnik}},
  \bibinfo{author}{\bibfnamefont{P.}~\bibnamefont{Dickinson}},
  \bibnamefont{et~al.}, \bibinfo{journal}{Phys. Rev. Lett.}
  \textbf{\bibinfo{volume}{70}}, \bibinfo{pages}{1553} (\bibinfo{year}{1993}).

\bibitem[{\citenamefont{Wollman et~al.}(1993)\citenamefont{Wollman,
  Van~Harlingen, Lee, Ginsberg, and Leggett}}]{Wollman93prl}
\bibinfo{author}{\bibfnamefont{D.~A.} \bibnamefont{Wollman}},
  \bibinfo{author}{\bibfnamefont{D.~J.} \bibnamefont{Van~Harlingen}},
  \bibinfo{author}{\bibfnamefont{W.~C.} \bibnamefont{Lee}},
  \bibinfo{author}{\bibfnamefont{D.~M.} \bibnamefont{Ginsberg}},
  \bibnamefont{and} \bibinfo{author}{\bibfnamefont{A.~J.}
  \bibnamefont{Leggett}}, \bibinfo{journal}{Phys. Rev. Lett.}
  \textbf{\bibinfo{volume}{71}}, \bibinfo{pages}{2134} (\bibinfo{year}{1993}).

\bibitem[{\citenamefont{Vekhter and Varma}(2003)}]{Vekhter03prl}
\bibinfo{author}{\bibfnamefont{I.}~\bibnamefont{Vekhter}} \bibnamefont{and}
  \bibinfo{author}{\bibfnamefont{C.~M.} \bibnamefont{Varma}},
  \bibinfo{journal}{Phys. Rev. Lett.} \textbf{\bibinfo{volume}{90}},
  \bibinfo{pages}{237003} (\bibinfo{year}{2003}).

\bibitem[{\citenamefont{Bok et~al.}(2010)\citenamefont{Bok, Yun, Choi, Zhang,
  Zhou, and Varma}}]{Bok10prb}
\bibinfo{author}{\bibfnamefont{J.~M.} \bibnamefont{Bok}},
  \bibinfo{author}{\bibfnamefont{J.~H.} \bibnamefont{Yun}},
  \bibinfo{author}{\bibfnamefont{H.-Y.} \bibnamefont{Choi}},
  \bibinfo{author}{\bibfnamefont{W.}~\bibnamefont{Zhang}},
  \bibinfo{author}{\bibfnamefont{X.~J.} \bibnamefont{Zhou}}, \bibnamefont{and}
  \bibinfo{author}{\bibfnamefont{C.~M.} \bibnamefont{Varma}},
  \bibinfo{journal}{Phys. Rev. B} \textbf{\bibinfo{volume}{81}},
  \bibinfo{pages}{174516} (\bibinfo{year}{2010}).

\bibitem[{\citenamefont{Sandvik et~al.}(2004)\citenamefont{Sandvik, Scalapino,
  and Bickers}}]{Sandvik04prb}
\bibinfo{author}{\bibfnamefont{A.~W.} \bibnamefont{Sandvik}},
  \bibinfo{author}{\bibfnamefont{D.~J.} \bibnamefont{Scalapino}},
  \bibnamefont{and} \bibinfo{author}{\bibfnamefont{N.~E.}
  \bibnamefont{Bickers}}, \bibinfo{journal}{Phys. Rev. B}
  \textbf{\bibinfo{volume}{69}}, \bibinfo{pages}{094523}
  (\bibinfo{year}{2004}).

\bibitem[{\citenamefont{Press et~al.}(2002)\citenamefont{Press, Teukolsky,
  Vetterling, and Flannery}}]{numericalrecipes}
\bibinfo{author}{\bibfnamefont{W.~H.} \bibnamefont{Press}},
  \bibinfo{author}{\bibfnamefont{S.~A.} \bibnamefont{Teukolsky}},
  \bibinfo{author}{\bibfnamefont{W.~T.} \bibnamefont{Vetterling}},
  \bibnamefont{and} \bibinfo{author}{\bibfnamefont{B.~P.}
  \bibnamefont{Flannery}}, \emph{\bibinfo{title}{Numerical recipes}}
  (\bibinfo{publisher}{Cambridge University Press, New York},
  \bibinfo{year}{2002}).

\bibitem[{\citenamefont{Aji et~al.}(2010)\citenamefont{Aji, Shekhter, and
  Varma}}]{Vivek10prb}
\bibinfo{author}{\bibfnamefont{V.}~\bibnamefont{Aji}},
  \bibinfo{author}{\bibfnamefont{A.}~\bibnamefont{Shekhter}}, \bibnamefont{and}
  \bibinfo{author}{\bibfnamefont{C.~M.} \bibnamefont{Varma}},
  \bibinfo{journal}{Phys. Rev. B} \textbf{\bibinfo{volume}{81}},
  \bibinfo{pages}{064515} (\bibinfo{year}{2010}).

\bibitem[{\citenamefont{Hwang et~al.}(2008)\citenamefont{Hwang, Schachinger,
  Carbotte, Gao, Tanner, and Timusk}}]{Hwang08prl}
\bibinfo{author}{\bibfnamefont{J.}~\bibnamefont{Hwang}},
  \bibinfo{author}{\bibfnamefont{E.}~\bibnamefont{Schachinger}},
  \bibinfo{author}{\bibfnamefont{J.~P.} \bibnamefont{Carbotte}},
  \bibinfo{author}{\bibfnamefont{F.}~\bibnamefont{Gao}},
  \bibinfo{author}{\bibfnamefont{D.~B.} \bibnamefont{Tanner}},
  \bibnamefont{and} \bibinfo{author}{\bibfnamefont{T.}~\bibnamefont{Timusk}},
  \bibinfo{journal}{Phys. Rev. Lett.} \textbf{\bibinfo{volume}{100}},
  \bibinfo{eid}{137005} (\bibinfo{year}{2008}).

\bibitem[{\citenamefont{Vignolle et~al.}(2007)\citenamefont{Vignolle, Hayden,
  McMorrow, Ronnow, Lake, Frost, and Perring}}]{Vignolle07naturephys}
\bibinfo{author}{\bibfnamefont{B.}~\bibnamefont{Vignolle}},
  \bibinfo{author}{\bibfnamefont{S.~M.} \bibnamefont{Hayden}},
  \bibinfo{author}{\bibfnamefont{D.~F.} \bibnamefont{McMorrow}},
  \bibinfo{author}{\bibfnamefont{H.~M.} \bibnamefont{Ronnow}},
  \bibinfo{author}{\bibfnamefont{B.}~\bibnamefont{Lake}},
  \bibinfo{author}{\bibfnamefont{C.~D.} \bibnamefont{Frost}}, \bibnamefont{and}
  \bibinfo{author}{\bibfnamefont{T.~G.} \bibnamefont{Perring}},
  \bibinfo{journal}{Naure Phys.} \textbf{\bibinfo{volume}{3}},
  \bibinfo{pages}{163} (\bibinfo{year}{2007}).

\bibitem[{\citenamefont{Zasadzinski et~al.}(2006)\citenamefont{Zasadzinski,
  Ozyuzer, Coffey, Gray, Hinks, and Kendziora}}]{Zasadzinski06prl}
\bibinfo{author}{\bibfnamefont{J.~F.} \bibnamefont{Zasadzinski}},
  \bibinfo{author}{\bibfnamefont{L.}~\bibnamefont{Ozyuzer}},
  \bibinfo{author}{\bibfnamefont{L.}~\bibnamefont{Coffey}},
  \bibinfo{author}{\bibfnamefont{K.~E.} \bibnamefont{Gray}},
  \bibinfo{author}{\bibfnamefont{D.~G.} \bibnamefont{Hinks}}, \bibnamefont{and}
  \bibinfo{author}{\bibfnamefont{C.}~\bibnamefont{Kendziora}},
  \bibinfo{journal}{Phys. Rev. Lett.} \textbf{\bibinfo{volume}{96}},
  \bibinfo{pages}{017004} (\bibinfo{year}{2006}).

\bibitem[{\citenamefont{Lanzara et~al.}(2001)\citenamefont{Lanzara, Bogdanov,
  Zhou, Kellar, Feng, Lu, Yoshida, Eisaki, Fujimori, Kishio
  et~al.}}]{Lanzara01nature}
\bibinfo{author}{\bibfnamefont{A.}~\bibnamefont{Lanzara}},
  \bibinfo{author}{\bibfnamefont{P.}~\bibnamefont{Bogdanov}},
  \bibinfo{author}{\bibfnamefont{X.}~\bibnamefont{Zhou}},
  \bibinfo{author}{\bibfnamefont{S.}~\bibnamefont{Kellar}},
  \bibinfo{author}{\bibfnamefont{D.}~\bibnamefont{Feng}},
  \bibinfo{author}{\bibfnamefont{E.}~\bibnamefont{Lu}},
  \bibinfo{author}{\bibfnamefont{T.}~\bibnamefont{Yoshida}},
  \bibinfo{author}{\bibfnamefont{H.}~\bibnamefont{Eisaki}},
  \bibinfo{author}{\bibfnamefont{A.}~\bibnamefont{Fujimori}},
  \bibinfo{author}{\bibfnamefont{K.}~\bibnamefont{Kishio}},
  \bibnamefont{et~al.}, \bibinfo{journal}{Nature}
  \textbf{\bibinfo{volume}{412}}, \bibinfo{pages}{510} (\bibinfo{year}{2001}).

\bibitem[{\citenamefont{Zhou et~al.}({2003})\citenamefont{Zhou, Yoshida,
  Lanzara, Bogdanov, Kellar, Shen, Yang, Ronning, Sasagawa, Kakeshita
  et~al.}}]{Zhou03nature}
\bibinfo{author}{\bibfnamefont{X.}~\bibnamefont{Zhou}},
  \bibinfo{author}{\bibfnamefont{T.}~\bibnamefont{Yoshida}},
  \bibinfo{author}{\bibfnamefont{A.}~\bibnamefont{Lanzara}},
  \bibinfo{author}{\bibfnamefont{P.}~\bibnamefont{Bogdanov}},
  \bibinfo{author}{\bibfnamefont{S.}~\bibnamefont{Kellar}},
  \bibinfo{author}{\bibfnamefont{K.}~\bibnamefont{Shen}},
  \bibinfo{author}{\bibfnamefont{W.}~\bibnamefont{Yang}},
  \bibinfo{author}{\bibfnamefont{F.}~\bibnamefont{Ronning}},
  \bibinfo{author}{\bibfnamefont{T.}~\bibnamefont{Sasagawa}},
  \bibinfo{author}{\bibfnamefont{T.}~\bibnamefont{Kakeshita}},
  \bibnamefont{et~al.}, \bibinfo{journal}{{Nature}}
  \textbf{\bibinfo{volume}{{423}}}, \bibinfo{pages}{{398}}
  (\bibinfo{year}{{2003}}).

\bibitem[{\citenamefont{Li et~al.}(2010)\citenamefont{Li, Baledent, Yu,
  Barisic, Hradil, Mole, Sidis, Steffens, Zhao, Bourges et~al.}}]{Li10nature}
\bibinfo{author}{\bibfnamefont{Y.}~\bibnamefont{Li}},
  \bibinfo{author}{\bibfnamefont{V.}~\bibnamefont{Baledent}},
  \bibinfo{author}{\bibfnamefont{G.}~\bibnamefont{Yu}},
  \bibinfo{author}{\bibfnamefont{N.}~\bibnamefont{Barisic}},
  \bibinfo{author}{\bibfnamefont{K.}~\bibnamefont{Hradil}},
  \bibinfo{author}{\bibfnamefont{R.~A.} \bibnamefont{Mole}},
  \bibinfo{author}{\bibfnamefont{Y.}~\bibnamefont{Sidis}},
  \bibinfo{author}{\bibfnamefont{P.}~\bibnamefont{Steffens}},
  \bibinfo{author}{\bibfnamefont{X.}~\bibnamefont{Zhao}},
  \bibinfo{author}{\bibfnamefont{P.}~\bibnamefont{Bourges}},
  \bibnamefont{et~al.}, \bibinfo{journal}{Nature}
  \textbf{\bibinfo{volume}{468}}, \bibinfo{pages}{283} (\bibinfo{year}{2010}).

\bibitem[{\citenamefont{He and Varma}(2011)}]{He11prl}
\bibinfo{author}{\bibfnamefont{Y.}~\bibnamefont{He}} \bibnamefont{and}
  \bibinfo{author}{\bibfnamefont{C.~M.} \bibnamefont{Varma}},
  \bibinfo{journal}{Phys. Rev. Lett.} \textbf{\bibinfo{volume}{106}},
  \bibinfo{pages}{147001} (\bibinfo{year}{2011}).

\end{thebibliography}

\end{document}